\documentclass{ws-ijmpa}

\usepackage[super,compress]{cite}
\usepackage{graphicx}
\usepackage{amsmath}
\begin{document}
	\markboth{Juhi Oudichhya}{Spectroscopic study of $D-$ meson in Regge phenomenology}
	
	\catchline{}{}{}{}{}
	%
	%%%%%%%%%%%%%%%%%%%%%%%%%%%%%%%%%%%%%%%%%%%%%%%%%%%%%%%%%%%%%%%%%%%%
	
	\title{Spectroscopic study of $D-$ meson in Regge phenomenology}

	\author{Juhi Oudichhya\footnote{
		Department of Physics, Sardar Vallabhbhai National Institute of Technology, Surat, Gujarat-395007, India}
	}
	
	\address{Department of Physics, Sardar Vallabhbhai National Institute of Technology,\\ Surat, Gujarat-395007, India
%		\footnote{
%			State completely without abbreviations, the affiliation and
%			mailing address, including country. Typeset in 8 pt italic.}
\\
		juhioudichhya01234@gmail.com}
	
	\author{Ajay Kumar Rai}
	
	\address{Department of Physics, Sardar Vallabhbhai National Institute of Technology, \\ Surat, Gujarat-395007, India \\
		raiajayk@gmail.com}
	
	\maketitle
	
	\begin{history}
		\received{Day Month Year}
		\revised{Day Month Year}
	\end{history}
	
	\begin{abstract}
		
		In last few years the experimental evidence of charmed mesons is increasing remarkably. In the present work we systematically studied the $D$  meson by employing Regge phenomenology.  By assuming the existence of the quasilinear Regge trajectories, several relations between Regge slope, intercept, and meson masses have been extracted. With the aid of these derived relations, Regge parameters are evaluated in both the ($J,M^{2}$) and ($n,M^{2}$) planes to obtain the mass spectra of $D$ meson.  In the forthcoming years, we believe that more candidates  will be reported and our  predictions could provide useful information for future experimental evidences.
	
	\end{abstract}
	
	\ccode{PACS numbers:}
	
	%\tableofcontents
	
	\section{Introduction}	
	In recent decades, the world wide experimental facilities have observed more and more higher excited singly heavy mesons particularly charmed mesons.
	The latest Particle Data Group (PDG) \cite{PDG} listed total 15 experimentally observed $D$ meson states with their quantum numbers, out of which 6 higher excited resonances still needs further confirmation. The $D^{*}(2640)^{\pm}$ and $D(3000)^{0}$ still awaiting the determination of their spin-parity quantum numbers. Till now only the low lying $1P$ and $1D$ states are observed experimentally. In last few years various experimental facilities such as Belle \cite{Belle2004}, BABAR \cite{AmoBABAR2010}, LHCb \cite{AaijLHCb2013,LHCb2,LHCb3,LHCb4}, FOCUS \cite{FOCUS}  etc. have observed many charmed mesons.  
	% have been focusing on searching for charmed mesons.
	In 2010, new resonances such as $D_{0}(2500)^{0}$, $D_{1}^{*}(2600)$, $D_{3}(2750)$, and $D_{1}^{*}(2760)$ were  firstly detected by the BABAR Collaboration \cite{AmoBABAR2010}, and further they were confirmed by LHCb Collaboration in 2013 with slightly different masses \cite{AaijLHCb2013}. Two new higher excited $D$ meson states; $D_{J}^{*}(3000)$ and $D_{J}(3000)$ were observed by LHCb Collaboration and further in 2016 another state, $D^{*}_{2}(3000)$ was observed with $J^{P}=2^{+}$ in the $D^{+} \pi^{-}$ channel. 
	
	The advancement of the experiments gives us excellent opportunity to establish a D-meson spectrum that is plentiful up to higher orbital and radial excitations.Theoretically, a lot of research has been done recently within various phenomenological approaches from various aspects to explain the nature of the charmed mesons, especially the newly detected states, and to establish the meson spectra .\cite{Chen2018,Devlani2013,Kher2017,Devlani2011,Chaturvedi2018,Chaturvedi2020,Rai2008,Matsuki2015,Rohit,AKRai2002}
	
	The authors of ref. \cite{Ebert2010} studies the heavy light mesons and obtained the mass spectra using the QCD-motivated relativistic quark model on the quasi potential approach. Further, the Regge trajectories have been drawn in the ($J,M^{2}$) and ($n,M^{2}$) plane which are linear as well as parallel. In a recent article \cite{Q. Li2022} another approach, a semi relativistic potential model is employed to evaluate the mass spectra of charmed mesons. Further the decay properties have also been studied. Cheung et. al. uses lattice QCD to study the highly excited charmonium, $D_{s}$, and $D$ meson spectra. In ref. \cite{Manan2016}, the authors computed the mass spectra of $D$ meson and also investigated their decay properties by employing the Martin like potential for the quark confinement. 
	
	In this work we compute the excited state masses of $D$ meson in the realm of Regge phenomenology. In our previous work we have used this approach to study the baryons ranging from light to singly, doubly, and triply heavy baryons 
	\cite{JuhiPRD1,JuhiPRD2,JuhiNPA,JuhiEPJA,physica,pramana}. 
	After the successful implementation of this theory on the investigation of baryons, now we use the same approach to study heavy light meson. From the quasi linear Regge trajectories, we have extracted some general relations having different flavors of mesons in terms of Regge slopes, intercepts, and meson masses. With the use of these derived relations, the Regge parameters have been calculated and further the mass spectra of charmed meson is computed in both the ($J,M^{2}$) and ($n,M^{2}$) plane.  
	
	The present article is structured as follows. After the briefing the newly observed experimental states and various theoretical approaches studies the charmed meson in Sec. 2 we explained the Regge theory and extracted the relations to calculate the higher excited states of $D$ mesons. In sec. 3 we thoroughly discussed our calculated results and their quantum numbers. Finally we concluded our work in Sec. 4.

\section{Theoretical Framework}

Regge theory examines almost all features of strong interactions, including spectra of particle, forces between the particles etc. According to the Regge theory, the hadrons are associated with Regge trajectories and the plots of mesons and baryons in the ($J,M^{2}$) plane, where $J$ is the total angular momentum and $M$ is the mass of the hadron, are usually called Chew Frautschi plots \cite{Chew1961}. They employed this theory to study the strong quark luon interaction and noticed that the experimental states of hadrons are found to be on linear trajectories. Hence, with the assumption of quasi linear Regge trajectories, one can write the general relation for Regge trajectories as \cite{JuhiPRD1,JuhiPRD2,Juhi_kaon2023,Wei2008},   
\begin{equation}
	\label{eq:1}
	J = \beta(0)+\beta^{'} M^2 ,
\end{equation}
where $\beta(0)$ and $\beta^{'}$ represent the intercept and slope of the trajectory, respectively. In this work, we take the following relations in terms of these Regge parameters for different flavors of meson multiplet, 
%	For a meson multiplet, the parameters for different quark constituents can be related by the following relations,
\begin{equation}
	\label{eq:2}
	\beta_{i \overline{i}}(0) + \beta_{j \overline{j}}(0) = 2a_{i \overline{j}}(0) ,	
\end{equation}

\begin{equation}
	\label{eq:3}
	\dfrac{1}{{\beta^{'}}_{i \overline{i}}} + \dfrac{1}{{\beta^{'}}_{j \overline{j}}} = \dfrac{2}{{\beta^{'}}_{i \overline{j}}} ,	
\end{equation}\\
where $i$ and $j$ represent quark flavours. Eq. (\ref{eq:2}), the additivity of intercepts was derived from the dual resonance model \cite{Kawarabayashi1969}, and was found to be satisfied in two dimensional QCD \cite{R. C. Brower1977},  and the quark bremsstrahlung model \cite{V.V. Dixit1979}. Eq. (\ref{eq:3}), the additivity of inverse slopes was derived in a model based on the topological expansion and the quark-antiquark string picture of hadrons \cite{A. B. Kaidalov1982}, it is also satisfied in the formal chiral limit and the heavy quark limit for both mesons and baryons \cite{L. Burakovsky1998}

Now by solving equations (\ref{eq:1}) and (\ref{eq:2}), we arrive to an another relation which is expressed as,  
\begin{equation}
	\label{eq:4}
	\beta^{'}_{i \overline{i}}M^{2}_{i \overline{i}}+\beta^{'}_{j \overline{j}}M^{2}_{j \overline{j}}=2\beta^{'}_{i \overline{j}}M^{2}_{i \overline{j}} .
\end{equation}
By combining the Eqs. (\ref{eq:3}) and (\ref{eq:4}) and solving the quadratic equation yield two pairs of solutions as, 
\begin{equation}
	\label{eq:5}
	\dfrac{\beta^{'}_{j \overline{j}}}{\beta^{'}_{i \overline{i}}}=\dfrac{1}{2M^{2}_{j \overline{j}}}\times[(4M^{2}_{i \overline{j}}-M^{2}_{i \overline{i}}-M^{2}_{j \overline{j}})
	\pm\sqrt{{{(4M^{2}_{i \overline{j}}-M^{2}_{i \overline{i}}-M^{2}_{j \overline{j}}})^2}-4M^{2}_{i \overline{i}}M^{2}_{j \overline{j}}}].
\end{equation}
and 
\begin{equation}
	\label{eq:6}	
	\dfrac{\beta^{'}_{i \overline{j}}}{\beta^{'}_{i \overline{i}}}=\dfrac{1}{4M^{2}_{i \overline{j}}}\times[(4M^{2}_{i \overline{j}}+M^{2}_{i \overline{i}}-M^{2}_{j \overline{j}})
	\pm\sqrt{{{(4M^{2}_{i \overline{j}}-M^{2}_{i \overline{i}}-M^{2}_{j \overline{j}}})^2}-4M^{2}_{i \overline{i}}M^{2}_{j \overline{j}}}].
\end{equation}
With the aid of Eq. (\ref{eq:5}) and the identical equation,
\begin{equation}
	\label{eq:7}
	\dfrac{\beta^{'}_{j \overline{j}}}{\beta^{'}_{i \overline{i}}}=	\dfrac{\beta^{'}_{k \overline{k}}}{\beta^{'}_{i \overline{i}}}\times	\dfrac{\beta^{'}_{j \overline{j}}}{\beta^{'}_{k \overline{k}}} .
\end{equation}
where $k$ can be any quark flavor, we have

\begin{eqnarray}
	\label{eq:8}
	%		\begin{align}
		\dfrac{[(4M^{2}_{i \overline{j}}-M^{2}_{i \overline{i}}-M^{2}_{j \overline{j}})+\sqrt{{{(4M^{2}_{i \overline{j}}-M^{2}_{i \overline{i}}-M^{2}_{j \overline{j}}})^2}-4M^{2}_{i \overline{i}}M^{2}_{j \overline{j}}}]}{2M^{2}_{j \overline{j}}} \\
		=\dfrac{[(4M^{2}_{i \overline{k}}-M^{2}_{i \overline{i}}-M^{2}_{k \overline{k}})+\sqrt{{{(4M^{2}_{i \overline{k}}-M^{2}_{i \overline{i}}-M^{2}_{k \overline{k}}})^2}-4M^{2}_{i \overline{i}}M^{2}_{k \overline{k}}}]/2M^{2}_{k \overline{k}}}{[(4M^{2}_{j \overline{k}}-M^{2}_{j \overline{j}}-M^{2}_{k \overline{k}})+\sqrt{{{(4M^{2}_{j \overline{k}}-M^{2}_{j \overline{j}}-M^{2}_{k \overline{k}}})^2}-4M^{2}_{j \overline{j}}M^{2}_{k \overline{k}}}]/2M^{2}_{k \overline{k}}} .
		%		\end{align}
\end{eqnarray}
This is the general relationship we have derived between meson masses of different flavors. The mass of any mesonic state can be calculated using this equation. 

\subsection{Masses of $D$ meson in the ($J,M^{2}$) plane}	
Hence, in this work we have used these extracted relations to determine the mass spectra of $D$ meson. To compute the excited state masses in the  ($J,M^{2}$) plane, the Regge parameters are determined. 
Due to the quark composition of $D$ meson which is composed of one light quark (u or d) and one charm quark ($c$),  we put $i=n (u$ or $d)$ and $j=c$ into Eq. (\ref{eq:6}), we get
%	(i) For $D$ meson ($n\overline{c}$): Put $i=n$ and $j=c$, we have
\begin{equation}
	\label{eq:9}
	\dfrac{\beta^{'}_{n \overline{c}}}{\beta^{'}_{n \overline{n}}}=\dfrac{1}{4M^{2}_{n \overline{c}}}\times[(4M^{2}_{n \overline{c}}+M^{2}_{n \overline{n}}-M^{2}_{c \overline{c}})
	\pm\sqrt{{{(4M^{2}_{n \overline{c}}-M^{2}_{n \overline{n}}-M^{2}_{c \overline{c}}})^2}-4M^{2}_{n \overline{n}}M^{2}_{c \overline{c}}}].
\end{equation}
\\
Therefore, after inserting the masses of $n \overline{n}$, $n \overline{c}$, and $c \overline{c}$ for $J^{P}$ = $0^{-}$ from PDG \cite{PDG}, we get the value of slope ratio, $\beta^{'}_{n \overline{c}}/\beta^{'}_{n \overline{n}}$. Now, from $1/M^{2}_{J+1}-M_{J}^{2}$, we can get $\beta^{'}_{n\overline{}n}$ = 0.6696$\pm$0.0035 GeV$^{-2}$. Thus we have $\beta^{'}_{n\overline{c}}$ = 0.4789$\pm$0.0025 GeV$^{-2}$ for $0^{-}$ trajectory. In the similar way we have evaluated  $\beta^{'}_{n\overline{c}}$ = 0.5700$\pm$0.0035 GeV$^{-2}$ for $1^{-}$ trajectory. 

Now, from Eq. (\ref{eq:1}) we can write,
\begin{equation}
	\label{eq:10}
	M_{J+1} = \sqrt{M_{J}^{2}+\dfrac{1}{\beta^{'}}}.
\end{equation}
Using the Eq. (\ref{eq:10}) and the of values of Regge slopes extracted for $D$ meson, the higher excited states ($L=1,2,3...$) lying on the $0^{-}$ and $1^{-}$ trajectories can be evaluated.
The calculated numerical values along with the experimental masses where available and the predictions of other theoretical approaches are displayed in Table \ref{tab:table1}.

%	\subsection{Subheadings}
%	
%	Subheadings should be typeset in boldface italic and capitalize
%	the first letter of the first word only. Section number to be in
%	boldface roman.

%	\section{Equations}
%	
%	Displayed equations should be numbered consecutively in each
%	section, with the number set flush right and enclosed in
%	parentheses
%	\begin{equation}
%		\mu(n, t) = \frac{\sum^\infty_{i=1} 1(d_i < t, N(d_i)
%			= n)}{\int^t_{\sigma=0} 1(N(\sigma) = n)d\sigma}\,.
%		\label{diseqn}
%	\end{equation}

%	\section{Figures}
%	
%	Figures are to be embedded in the text nearest their first reference
%	and sequentially numbered in Arabic numerals. The caption must be
%	placed below the figure (see Fig.~\ref{f1}) and typeset in 8 pt roman with
%	baselineskip of 10 pt. Use double spacing between a caption and the
%	text that follows immediately.
%	
%	\begin{figure}[b]
%		\centerline{\includegraphics[width=3.8cm]{ijmpaf1}}
%		\caption{A schematic illustration of dissociative recombination. The
%			direct mechanism, 4m$^2_\pi$ is initiated when the
%			molecular ion S$_{\rm L}$ captures an electron with
%			kinetic energy. \label{f1}}
%	\end{figure}
%	

%\section{Tables}
%		
	\begin{table}[h]
		\tbl{Masses of excited states of the $D$ meson in the $(J,M^{2})$ plane (in MeV).}
%		\label{tab:table1}
		{\begin{tabular}{@{}cccccccccccccccc@{}} \toprule
				\textit{$N^{2S+1}L_{J}$}&Present& PDG \cite{PDG} & \cite{Godfrey2016} & \cite{Ebert2010} & \cite{Q. Li2022}&\cite{Vpatel}  	\\
			 \colrule
			$1^{1}S_{0}$ & 1864.84$\pm$0.05 & 1864.84$\pm$0.05 & 1877 & 1871 & 1865 &1889\\ 
		$1^{1}P_{1}$ & 2359.18$\pm$2.31 & 2422.1$\pm$0.6 & 2467 & 2426 & 2453 & 2448\\ 
		$1^{1}D_{2}$ & 2766.56$\pm$2.78 & 2747$\pm$6 & 2845 & 2806 &2827 &2754\\
		$1^{1}F_{3}$ & 3121.22$\pm$3.02 & &3143 & 3129 & 3129&3051\\
		$1^{1}G_{4}$ & 3439.49$\pm$3.16 & &3399 & 3403\\
		$1^{1}H_{5}$ & 3730.71$\pm$3.26 & &\\ 
		\noalign{\smallskip}  
		$1^{3}S_{1}$ & 2006.85$\pm$0.05 &2006.85$\pm$0.05 & 2041 & 2010 & 2008 &2007 \\ 
		$1^{3}P_{2}$ & 2404.54$\pm$2.24 & 2461.1$^{+0.7}_{-0.8}$ & 2502 & 2460 & 2475 &2462\\ 
		$1^{3}D_{3}$ & 2745.22$\pm$2.77 & 2763.1$\pm$3.2 & 2833 & 2863 & 2782 &2807\\  
		$1^{3}F_{4}$ & 3048.05$\pm$3.06 & &3113 & 3187 & 3034 &3029\\ 
		$1^{3}G_{5}$ & 3323.40$\pm$3.24 & &3362 &3473\\
		$1^{3}H_{6}$ & 3577.62$\pm$3.36 & &\\
			 \botrule
			\end{tabular} 	\label{tab:table1}}
	\end{table}

%With the help of $\alpha^{'}_{(S)}$ and $a_{0(S)}$, we can evaluate the masses of the excited $D$-meson states for $n$ = 3, 4, 5... Similarly, we can express these relations for $P$ and $D$-wave as,
%\begin{equation}
%	\label{eq:13}
%	\begin{split}
%		1 = \beta_{0(P)} + \beta_{(P)} M^{2}_{n\overline{c}(1P)},\\
%		2 = \beta_{0(P)} + \beta_{(P)} M^{2}_{n\overline{c}(2P)},\\
%		1 = \beta_{0(D)} + \beta_{(D)} M^{2}_{n\overline{c}(1D)},\\
%		2 = \beta_{0(D)} + \beta_{(D)} M^{2}_{n\overline{c}(2D)},
%	\end{split}
%\end{equation}
%With the help of above obtained relations we have extracted the values of 
%Regge slopes and intercepts for D mesons for each Regge trajectory with spin $s$= 0 and 1, and obtain the radially excited state masses lying on that trajectory. The predicted mass spectra for D meson is shown in Table 2 along with the predictions of other theoretical approaches. 
	\subsection{Masses of $D$ meson in the ($n,M^{2}$) plane}
After obtaining the orbitally excited state masses of $D$ mesons, this section is dedicated to calculate the radial excitations and for that the Regge slopes and intercepts have been extracted. 
The general equation for linear Regge trajectories in the ($n,M^{2}$) plane can be expressed as,
\begin{equation}
	\label{eq:11}
	n = a_{0} + \alpha^{'} M^{2},
\end{equation}
where $n$ represents the radial principal quantum number; 1,2,3.... $a_{0}$, and $\alpha^{'}$ are the intercept and slope of the trajectories. 
The Regge slope ($\alpha^{'}$) and Regge intercept ($a_{0}$) are the same for the meson multiplets lying on the single Regge trajectory.
To determine the radially excited states of $D$ meson, the Regge parameters have been extracted for $S$, $P$, $D$... states using the Eq. (\ref{eq:11}).  
%Using relation (\ref{eq:11}), the values of $\beta$ and $\beta_{0}$ for D meson can be extracted for $S$, $P$, $D$... states to evaluate the excited state masses lying on Regge trajectories.
From the slope equation, we have $\alpha^{'}_{(S)} = 1/(M^{2}_{n\overline{c}(2S)}-M^{2}_{n\overline{c}(1S)})$. For the computation, the masses of first radial excitations ($n$=2) are taken as inputs from PDG \cite{PDG}, the experimentally observed masses wherever available. While, due to the lack of experimental availability in some states we have taken the theoretical predictions of \cite{Ebert2010}.
%Here the mass values of first radial excitations ($n$=2)  are taken as inputs from PDG \cite{PDG}, the experimentally observed masses wherever available. Whereas, due to the unavailability of experimental masses in some states we have taken the theoretical predictions of \cite{Ebert2009} for the calculation. 
%Here also, the error analysis is incorporated for the calculation of Regge parameters. 
%The detailed calculation to incorporate the experimental errors in  $\beta$ and $\beta_{0}$ is given in appendix.
By inserting the values of  $M_{n\overline{c}(1S)}$ and $M_{n\overline{c}(2S)}$ for $J^{P}=0^{-}$, we can get $\alpha^{'}_{(S)}$ = 0.33115$\pm$0.011 GeV$^{-2}$ for $S$-states with spin $s$=0. Now from relation (\ref{eq:11}) we can write, 

\begin{equation}
	\label{eq:12}
	\begin{split}
		1 = a_{0(S)} + \alpha^{'}_{(S)} M^{2}_{n\overline{c}(1S)},\\
		2 = a_{0(S)} + \alpha^{'}_{(S)} M^{2}_{n\overline{c}(2S)},
	\end{split}
\end{equation}
Solving the above equations, we get $a_{0(S)}$ = -0.15162$\pm$0.037. In the same way the values of Regge slopes and intercept can be extracted for $P$ and $D$ states also for spin $s=0$ and 1. Hence, with the aid of these Regge parameters the radially excited states of $D$-meson for $n$ = 3, 4, 5... can be evaluated. The predicted mass spectra is shown in Table \ref{tab:table2} along with the predictions of other theoretical approaches.

	\begin{table}[h]
	\tbl{Masses of excited states of the $D$ meson in the $(n,M^{2})$ plane (in MeV).}
	{\begin{tabular}{@{}cccccccccccc@{}} \toprule
			\textit{$N^{2S+1}L_{J}$}&Present& \cite{Godfrey2016} & \cite{Q. Li2022} & \cite{Devlani2013} & \cite{Kher2017}	\\  
			\colrule
		$1^{1}S_{0}$ & 1864.84$\pm$0.05 & 1877 &1865 &1865 &1884 \\ 
	$2^{1}S_{0}$ & \textbf{2549.0$\pm$19.00} \cite{PDG} &2581 & 2547 & 2598 & 2582\\ 
	$3^{1}S_{0}$ & 3048.99$\pm$52.55 &3068 & 3029 &3087 &3186\\ 
	$4^{1}S_{0}$ & 3540.75$\pm$58.80 &3468 & & 3498 & 3746\\ 
	$5^{1}S_{0}$ & 3944.20$\pm$64.68& 3814\\ 
	$6^{1}S_{0}$ & 4310.04$\pm$70.17\\ 
	\noalign{\smallskip}
	$1^{3}S_{1}$ & 2006.85$\pm$0.05 & 2041 & 2008 & 2018 & 2010 \\ 
	$2^{3}S_{1}$ & \textbf{2627.0$\pm$10.00} \cite{PDG}& 2643 & 2636 & 2639 & 2655\\ 
	$3^{3}S_{1}$ & 3126.47$\pm$30.89 & 3110 & 3093 & 3110 & 3239\\ 
	$4^{3}S_{1}$ & 3556.47$\pm$34.11 &3497 & & 3514 & 3789\\ 
	$5^{3}S_{1}$ & 3939.82$\pm$37.19 & 3837\\ 
	$6^{3}S_{1}$ & 4289.04$\pm$40.12\\ 
	
	\hline
	$1^{1}P_{1}$ &2359.18$\pm$2.31 & 2456 & 2453 & 2454 & 2447\\
	$2^{1}P_{1}$ &\textbf{2932.0$\pm$0.00} \cite{Ebert2010} & 2924 & 2936 & 2951 & 3034\\
	$3^{1}P_{1}$ &3409.91$\pm$6.94& 3328 & & &3582\\
	$4^{1}P_{1}$ &3828.63$\pm$7.46 &3681 \\
	$5^{1}P_{1}$ &4205.86$\pm$7.99 \\  
	\noalign{\smallskip}
	$1^{3}P_{2}$ & 2404.54$\pm$2.24 & 2502 & 2475 & 2473 & 2461 \\
	$2^{3}P_{2}$ & \textbf{3012.0$\pm$0.00}  \cite{Ebert2010}& 2957 & 2955 & 2971 & 3039\\
	$3^{3}P_{2}$ & 3516.03$\pm$6.50 & 3353 & & &3584\\
	$4^{3}P_{2}$ & 3956.36$\pm$7.00 &3701\\
	$5^{3}P_{2}$ & 4352.37$\pm$7.51\\ 
	\hline
	$1^{1}D_{2}$ & 2766.56$\pm$2.78 & 2816 & 2827 & 2829 & 2783\\ 
	$2^{1}D_{2}$ & \textbf{3259.0$\pm$0.00}  \cite{Ebert2010} & 3212 & 3221 & 3256 & 3341\\ 
	$3^{1}D_{2}$ & 3686.28$\pm$10.87 &3566 & & & 3873\\ 
	$4^{1}D_{2}$ & 4068.93$\pm$11.46\\ 
	$5^{1}D_{2}$ & 4418.57$\pm$12.96\\ 
	\noalign{\smallskip}
	$1^{3}D_{3}$ & 2745.22$\pm$2.77 & 2833 & 2782 & 2741 & 2788\\
	$2^{3}D_{3}$ & \textbf{3351.0$\pm$0.00} \cite{Ebert2010} & 3226 & 3202 & 3187 & 3355\\ 
	$3^{3}D_{3}$ & 3835.13$\pm$9.32&3579 & & &3885\\
	$4^{3}D_{3}$ & 4277.17$\pm$9.94\\
	$5^{3}D_{3}$ & 4677.63$\pm$10.58\\ 
			
			\botrule
		\end{tabular} 	\label{tab:table2}}
\end{table}

\begin{figure}
%\centering	
\hspace{-0.8cm}
	\begin{minipage}{0.5\textwidth}
		
		\includegraphics[scale=0.2]{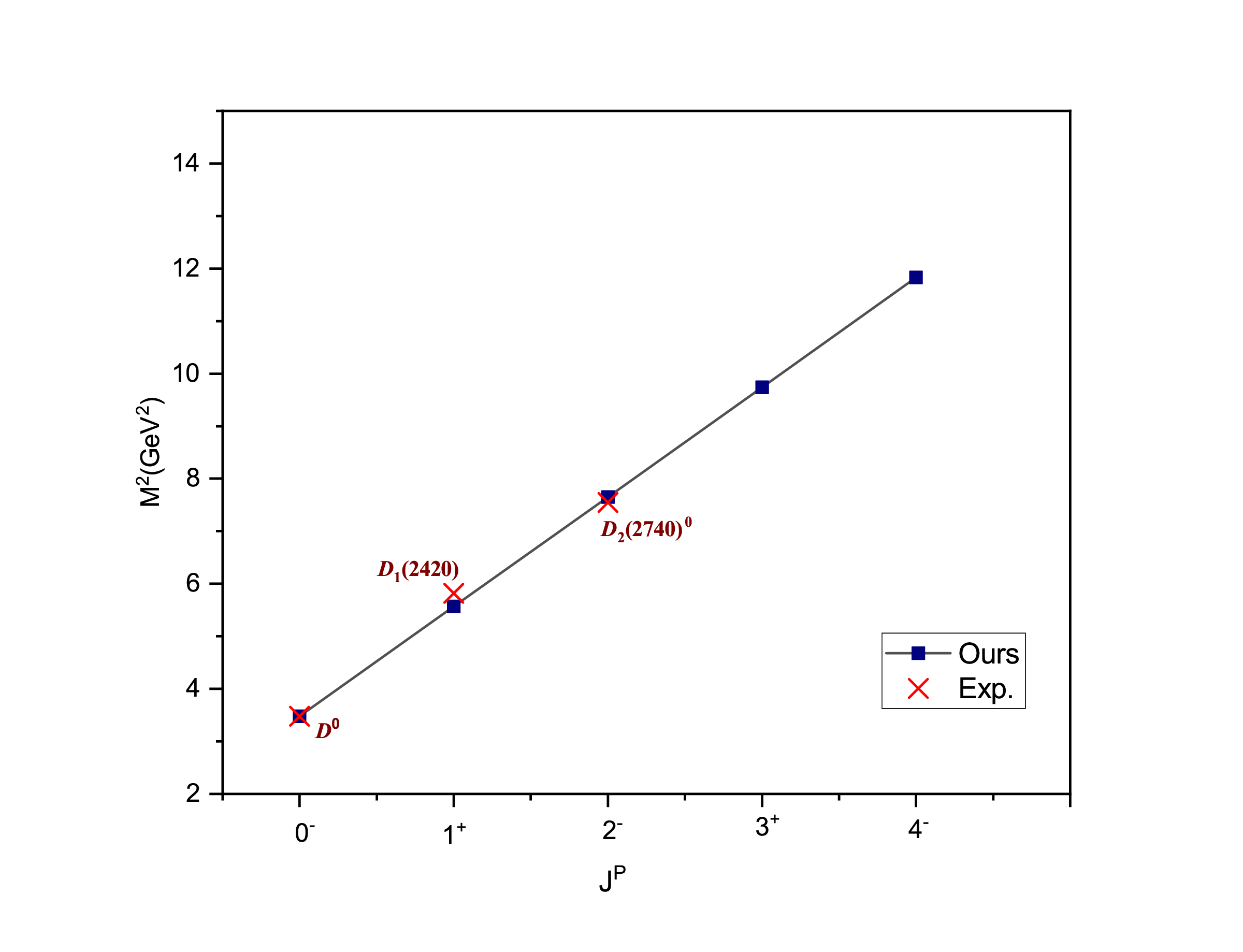}
		\caption{\label{fig:1}{Regge trajectory in the ($J,M^{2}$) plane for $D$ meson with unnatural parity states}}
	\end{minipage}
	\begin{minipage}{0.5\textwidth}
	\hspace{-1 cm}
%\centering
		\includegraphics[scale=0.2]{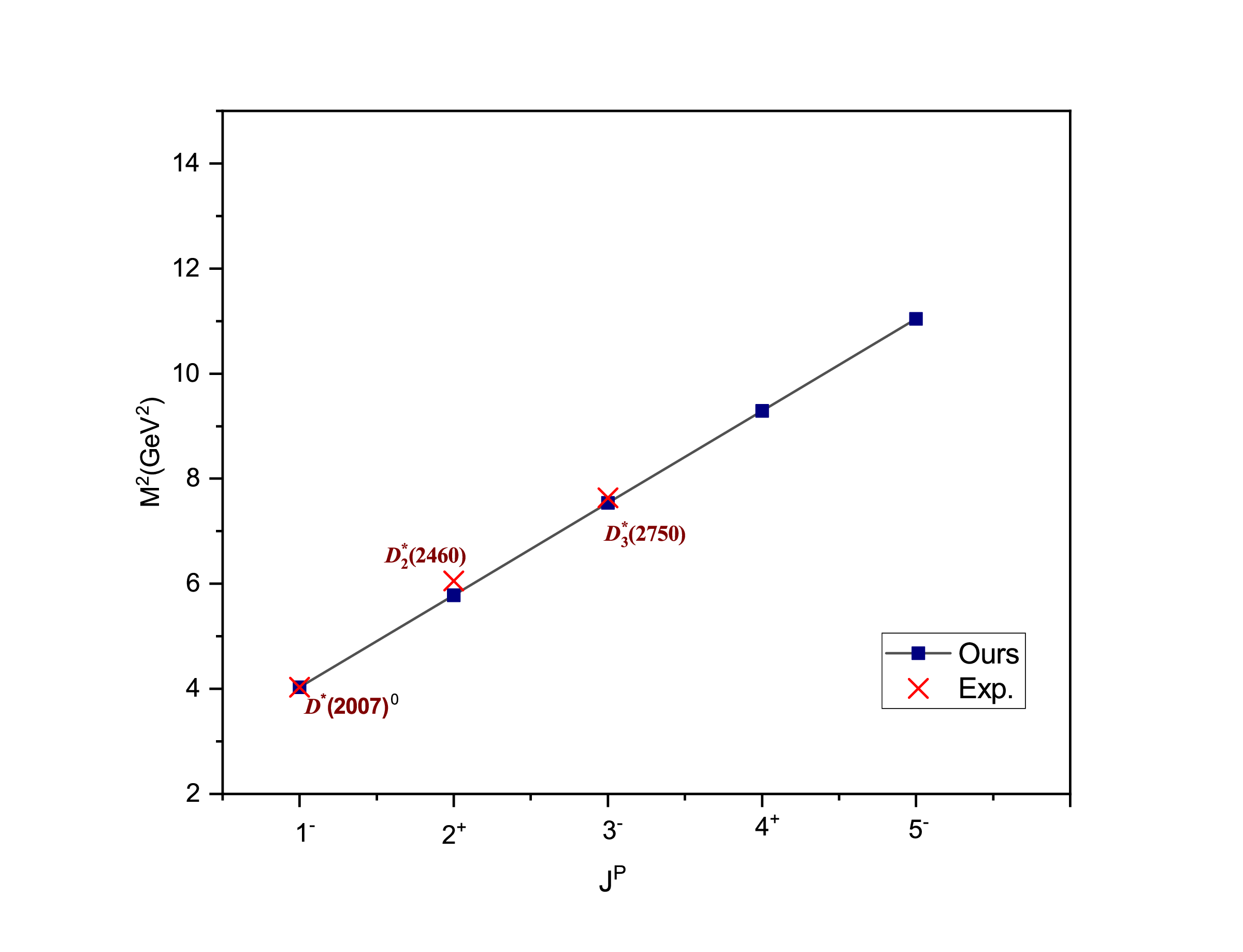}
		\caption{\label{fig:2}{Regge trajectory in the ($J,M^{2}$) plane for $D$ meson with natural parity states}}
	\end{minipage}
	
\end{figure}

\begin{figure}
	
	\hspace{-0.8 cm}	
	\begin{minipage}{0.5\textwidth}
	
%\centering
		\includegraphics[scale=0.2]{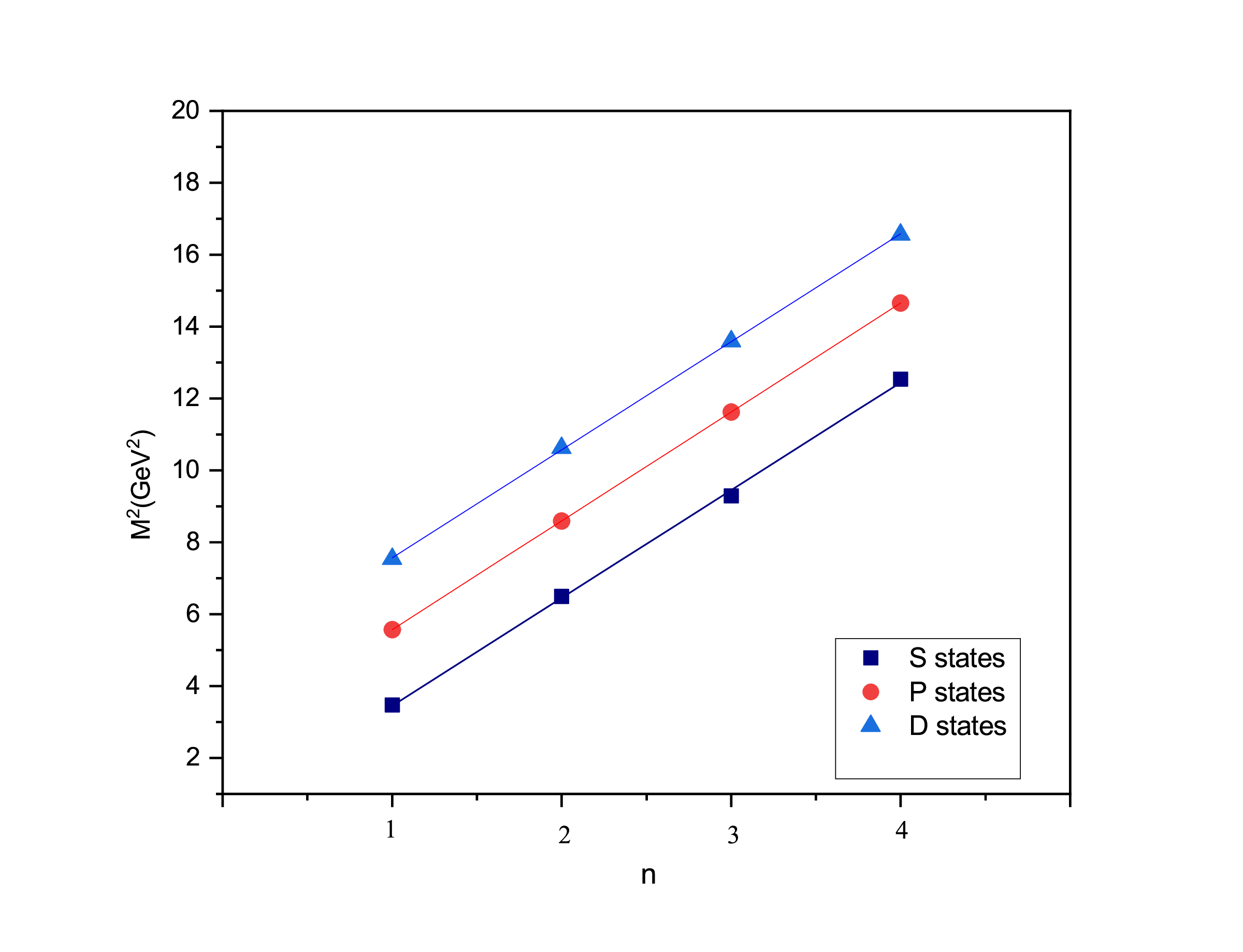}
		\caption{\label{fig:3}{Regge trajectory in the ($n,M^{2}$) plane for $D$ meson with unnatural parity states}}
	\end{minipage}
	\begin{minipage}{0.5\textwidth}
%		\hspace{-1 cm}
%\centering
		\includegraphics[scale=0.2]{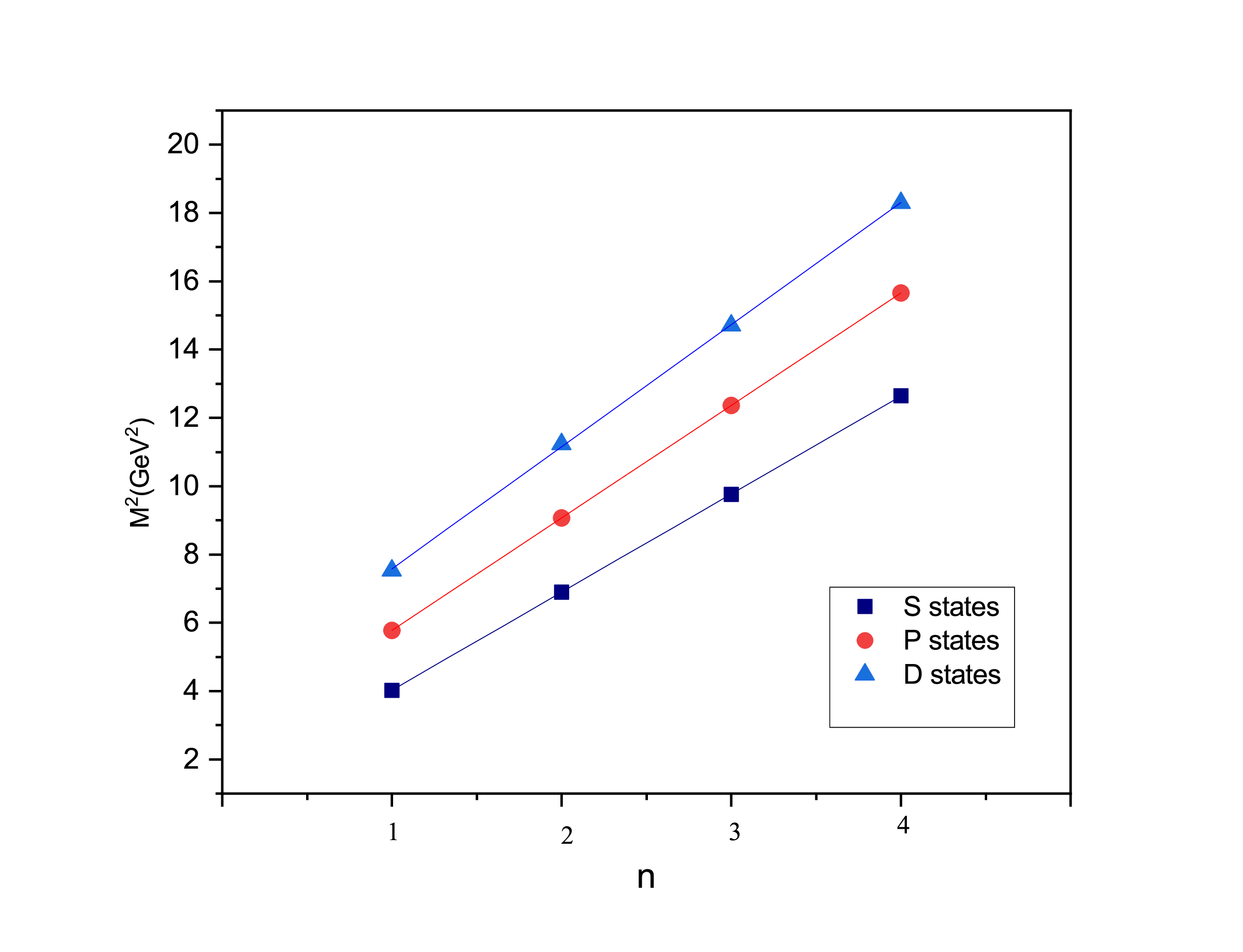}
		\caption{\label{fig:4}{Regge trajectory in the ($n,M^{2}$) plane for $D$ meson with natural parity states}}
	\end{minipage}
\end{figure}

\section{Results and Discussion}

The present article is dedicated to the systematic study of charmed meson including the determination of mass spectra and assignment of spin parity quantum numbers using Regge theory. In $D$ meson family the ground state masses are well established and in the present work we have taken these experimental ground state masses as inputs to extract the Regge parameters and to evaluate the higher excited state masses using the relations we have derived.
The calculated masses of orbital and radial excitation of $D$ meson in the ($J,M^{2}$) and ($n,M^{2}$) planes are presented in the tables \ref{tab:table1} and \ref{tab:table2} respectively.  Further, the Regge trajectories are also constructed in both the planes. Our obtained masses are presented on the Regge line along with the experimentally detected masses which are presented by cross sign as shwon in figs. 1-4. 

We compared our calculated outcomes with the available experimental masses and also with the predictions of other theoretical models. 
In the latest PDG \cite{PDG} the experimentally observed states $D_{1}(2420)$ and $D_{2}^{*}(2460)$ having masses 2422.1$\pm$0.6 MeV and 2461.1$^{+0.7}_{-0.8}$ MeV respectively. Both these states are assigned with positive parity having $J^{P}$ = $1^{+}$ and $2^{+}$ respectively. Our calculated masses for $1^{1}P_{1}$ and $1^{3}P_{2}$ states are 2359.18$\pm$2.31 MeV and 2404.54$\pm$2.24 MeV respectively which are slightly lower than the experimentally observed masses with a mass difference of 57-63 MeV. But these two states; $D_{1}(2420)$ and $D_{2}^{*}(2460)$ can be assigned as $1^{1}P_{1}$ and $1^{3}P_{2}$ respectively. Further, in $D$ wave our predicted mass for the low lying $1^{1}D_{2}$ state is 2766.56$\pm$2.78 MeV which is very close to the experimentally observed resonance $D_{2}(2740)^{0}$ mentioned in PDG having mass 2747$\pm$6 MeV and $J^{P}$ value $2^{-}$. The mass difference between our obtained mass and the experimental mass is only 19 MeV. Hence, this state can be assigned as $1^{1}D_{2}$ state. One more state listed in PDG \cite{PDG} with negative parity $3^{-}$ is $D_{3}^{*}(2750)$ having experimental mass 2763.1$\pm$3.2 MeV is found to be very close to our calculated mass of $1^{3}D_{3}$ state as 2745.22$\pm$2.77 MeV having a mass difference of 18 MeV. Hence, we can assign this state to be $1^{3}P_{2}$. 

Apart from this we compared our outcomes with the results of other theoretical models as well. A variety of predictions can be seen from the different approaches. Our results shows a consistent behavior with the outcomes of refs. \cite{Ebert2010,Godfrey2016,Q. Li2022} with a mass difference of few MeV. Some higher excited states shows slight discrepancies with the results of other theoretical models.  Further the Regge trajectories are also constructed in both the ($J,M^{2}$) and $(n,M^{2})$ planes as shown in Figs 1-4. We plotted our obtained data lying on straight line along with the experimentally available masses present by cross sign. It can be shown that the experimental resonances lies very close to our predicted masses. 

\section{Conclusion}

The mass spectra of charmed meson is calculated in this work. The computed outcomes are consistent with the experimental masses as well as shows a general agreement with the predictions of other theoretical models. Various phenomenological approaches gives a variety of results, hence more computation and study of charmed meson is required for the identification of higher states of charmed mesons. Our large number of mass value predictions and the assignments of their spin-parity quantum numbers will provide useful information in various experimental groups such as BABAR, LHCb, Belle etc. and the upcoming experimental facility PANDA \cite{PANDA1,PANDA2,PANDA3} to explore the higher excited states.    

\section{Acknowledgments}  
The authors are thankful to the organisers of 11$^{th}$ international conference on new frontiers in physics (ICNFP
2022) for providing the opportunity to present our work.

%	

%	\section*{Acknowledgments}
%	
%	This section should come before the References. Dedications and funding
%	information may also be included here.
%	
	\appendix
	
%	\section{Appendices}
%	
%	Appendices should be used only when absolutely necessary. They
%	should come before the References. If there is more than one
%	appendix, number them alphabetically. Number displayed equations
%	occurring in the Appendix in this way, e.g.~(\ref{app1}),\break
%	(\ref{app2}), etc.

	%\begin{thebibliography}{000} %for 3 digits
	%\begin{thebibliography}{00}  %for 2 digits
	
\end{document}